%% file: main.tex
\documentclass[aps,prapplied,reprint,showpacs,amsmath,amssymb,floatfix,superscriptaddress,noeprint]{revtex4-2}
\usepackage{bm}
\usepackage{color}
\usepackage[colorlinks,linkcolor=blue,urlcolor=blue,citecolor=blue,anchorcolor=blue]{hyperref}

\usepackage{graphicx}
\usepackage[capitalise]{cleveref}
\usepackage{physics}
\usepackage[detect-weight=true, separate-uncertainty=true]{siunitx}
\usepackage{multirow}
\usepackage{circledsteps}
\usepackage{xcolor}
\usepackage{ulem}
\begin{document}

\title{Transmon-qubit readout using in-situ bifurcation amplification in the mesoscopic regime}

\author{R. Dassonneville}
\affiliation{Univ. Grenoble-Alpes, CNRS, Grenoble INP, Institut N\'eel, 38000 Grenoble, France}
\affiliation{Aix Marseille Univ., CNRS, IM2NP, Marseille, France}
\author{T. Ramos}
\affiliation{Institute of Fundamental Physics, IFF-CSIC, Calle Serrano 113b, 28006 Madrid, Spain}
\author{V. Milchakov}
\affiliation{Univ. Grenoble-Alpes, CNRS, Grenoble INP, Institut N\'eel, 38000 Grenoble, France}
\author{C. Mori}
\affiliation{Univ. Grenoble-Alpes, CNRS, Grenoble INP, Institut N\'eel, 38000 Grenoble, France}
\author{L. Planat}
\affiliation{Univ. Grenoble-Alpes, CNRS, Grenoble INP, Institut N\'eel, 38000 Grenoble, France}
\author{F. Foroughi}
\affiliation{Univ. Grenoble-Alpes, CNRS, Grenoble INP, Institut N\'eel, 38000 Grenoble, France}
\author{C. Naud}
\affiliation{Univ. Grenoble-Alpes, CNRS, Grenoble INP, Institut N\'eel, 38000 Grenoble, France}
\author{W. Hasch-Guichard}
\affiliation{Univ. Grenoble-Alpes, CNRS, Grenoble INP, Institut N\'eel, 38000 Grenoble, France}
\author{J. J. Garc{\'i}a-Ripoll}
\affiliation{Institute of Fundamental Physics, IFF-CSIC, Calle Serrano 113b, 28006 Madrid, Spain}
\author{N. Roch}
\affiliation{Univ. Grenoble-Alpes, CNRS, Grenoble INP, Institut N\'eel, 38000 Grenoble, France}
\author{O. Buisson}
\affiliation{Univ. Grenoble-Alpes, CNRS, Grenoble INP, Institut N\'eel, 38000 Grenoble, France}

\date{\today}
\begin{abstract}
We demonstrate a transmon qubit readout based on the nonlinear response to a drive of polaritonic meters in-situ coupled to the qubit. Inside a 3D readout cavity, we place a transmon molecule consisting of a transmon qubit and an ancilla mode interacting via non-perturbative cross-Kerr coupling. The cavity couples strongly only to the ancilla mode, leading to hybridized lower and upper polaritonic meters. Both polaritons are anharmonic and dissipative, as they inherit a self-Kerr nonlinearity $U$ from the ancilla and effective decay $\kappa$ from the open cavity. Via the ancilla, the polariton meters also inherit the non-perturbative cross-Kerr coupling to the qubit. This results in a high qubit-dependent displacement $2\chi > \kappa, ~U$ that can be read out via the cavity without causing Purcell decay. Moreover, the polariton meters, being nonlinear resonators, present bistability, and bifurcation behavior when the probing power increases. In this work, we focus on the bifurcation at low power in the few-photon regime, called the mesoscopic regime, which is accessible when the self-Kerr and decay rates of the polariton meter are similar $U\sim \kappa$. Capitalizing on a latching mechanism by bifurcation, the readout is sensitive to transmon qubit relaxation error only in the first tens of nanoseconds. We thus report a single-shot fidelity of \SI{98.6}{\%} while having an integration time of a \SI{500}{ns} and no requirement for an external quantum-limited amplifier.
    
\end{abstract}

\maketitle
\section{Introduction}
 Qubit state readout is a mandatory step in quantum information processing. For superconducting circuits, the dispersive readout is the standard scheme \cite{PRApplied_Walter_2017, Sunada2022}. It relies on the transverse interaction between an anharmonic mode, whose first two levels are used as a qubit, and another mode, usually harmonic, used as a meter \cite{Blais_PRA_2004,PRA_Koch_2007}. This transverse interaction couples the qubit polarization to the meter field quadrature and hybridizes the qubit with the meter. In perturbation theory, the resulting dispersive interaction (or perturbative cross-Kerr coupling) corresponds in first order to an energy-energy interaction where the qubit state shifts the meter frequency and reciprocally, the number of photons in the meter shifts the qubit frequency. When applying a coherent pulse close to the meter frequency for a time smaller than the relaxation time of the qubit $T_1$, the qubit state is inferred by distinguishing, in phase-space of the acquired output field, the two pointer states of the meter corresponding to the qubit excited or ground states. 
 Using this dispersive readout, single-shot readout with high fidelity is nowadays routinely achieved, notably thanks to quantum-limited Josephson Parametric Amplifier (JPA) \cite{Aumentado2020}. 
 However, the dispersive interaction contains intrinsic limitations, due to the higher order corrections in perturbation theory. The qubit states are slightly dressed by the meter states, which leads to Purcell decay \cite{Houck2008} and prevents from an ideal quantum non-demolition (QND) readout \cite{Pereira_PRL_2022, pereira_parallel_2023}. In addition, unwanted effects for the readout such as relaxation and excitation rate of the qubit can increase with readout photon number $\overline{n}$ \cite{Johnson2011,Minev2019,Lescanne2019,khezri2022measurementinduced,Shillito2022}. To overcome these limitations, a non-perturbative cross-Kerr coupling between the qubit and the meter has been proposed \cite{PRA_Diniz_2013,Wang2019} and demonstrated thanks to the property of a transmon molecule \cite{Dumur2015,Roy2017,Roy2018,Dassonneville2020} achieving high fidelity and QND single shot readout of a transmon qubit \cite{Dassonneville2020}. This result was realized through a polariton meter in its linear regime, whose signal was amplified through an external JPA.

 Alternatively to JPA, superconducting qubit readout can also be performed using a Josephson Bifurcation Amplifier (JBA) \cite{Lupascu2006,Vijay2009, mallet_single-shot_2009}. The JBA is a nonlinear pumped resonator such as the JPA, but it is pumped at different working point, where it presents a nonlinear amplification relationship between its input amplitude and output amplitude, leading to two stable states of small and large output amplitude for input signal below and above the bifurcation threshold, respectively. The information on the qubit state is then encoded into those two output states. In addition, the bifurcation presents hysteresys leading to a latching readout. The JBA dynamics is controlled by the detuning between the nonlinear resonator and the pump, the resonator losses $\kappa$ and its anharmonicity $U$. A same-chip implementation allows a direct coupling between the qubit and the JBA, with an in-situ amplifying bifurcation, greatly increasing the quantum detection efficiency \cite{Siddiqi2006,Lupascu2006,Boulant2007,mallet_single-shot_2009,Dewes2012,Eddins2019,Rosenthal2021}.  Here, we propose a readout based on bifurcation amplification of the in-situ nonlinear polariton meter.
 
 Up to now, the bifurcation readout has been realized in the weak anharmonicity limit $U \ll \kappa$ in which the bistability regime is achieved when the photon number $\overline{n}$ in the nonlinear resonator exceeds the critical number $N_\mathrm{crit}=\kappa/(3\sqrt{3}U) \gg 1$ \cite{Ong2011,Ong2013}. 
 However, this large photon number, needed to reach the bistability and thus bifurcation, exposes the qubit to excess backaction of the nonlinear cavity \cite{Nakano2009,Laflamme2012,Boissonneault2012,Ong2011,Ong2013} like inducing qubit state transitions and thus rendering the readout not QND. Instead of the usual classical regime of weak anharmonicity and large photon number ($U \ll \kappa$, $N_\mathrm{crit} \gg 1$), the excess backaction on the qubit could be weakened in the mesoscopic regime ($U \sim \kappa$, $N_\mathrm{crit} \sim 1$) where bistability appears with photon number close to unity $\overline{n} \gtrsim N_\mathrm{crit} \sim 1$. This little explored regime is in between the classical regime and the quantum regime ($U \gg \kappa$, $N_\mathrm{crit} \ll 1$, where the system behaves as an effective quantum few-level-system) \cite{Claudon2008, Muppalla2018, Andersen2020}.  
 
 In this paper, we demonstrate a transmon qubit state latching readout using an in-situ bifurcation of a polariton resonator in the mesoscopic regime. The polaritons are superpositions of the harmonic cavity mode and the anharmonic ancillary mode of the transmon molecule. They result from the strong coupling/hybridization between them. Both polariton modes inherit the nonlinearity of the ancilla mode, so they effectively behave as two nonlinear Kerr resonators exhibiting bistability \cite{Sarchi2008,Eichler2014,Winkel2020, Fischer2021}. By adjusting the ancilla frequency through an external magnetic flux, we control the hybridization, and consequently the anharmonicity and dissipation of each polariton. In \cref{sec:hybridization}, we discuss details of the transmon molecule, the ancilla-cavity hybridization, the resulting polariton modes and their tunability. 
 Contrary to our previous work \cite{Dassonneville2020} where we considered the linear regime with $\overline{n} \ll N_\mathrm{crit}$ and $N_\mathrm{crit} \gg 1$, here we investigate the nonlinear regime of the polaritons at large occupation $\overline{n} \gtrsim N_\mathrm{crit} \sim 1$. The polaritons response to a strong drive and its dependence on the qubit state are detailed in \cref{sec:transmission}. The hysteretic bistability behavior of the nonlinear upper polaritonic meter is analyzed in \cref{sec:bistability}. Finally, in \cref{sec:readout}, we take advantage of this bistability to perform a latching readout of the qubit state with a high single-shot fidelity and without any external quantum-limited amplifier.

\section{Tuning the ancilla-cavity hybridization}
\label{sec:hybridization}
\subsection{Transmon molecule in a cavity}

\begin{figure}
\includegraphics[width=8.6cm]{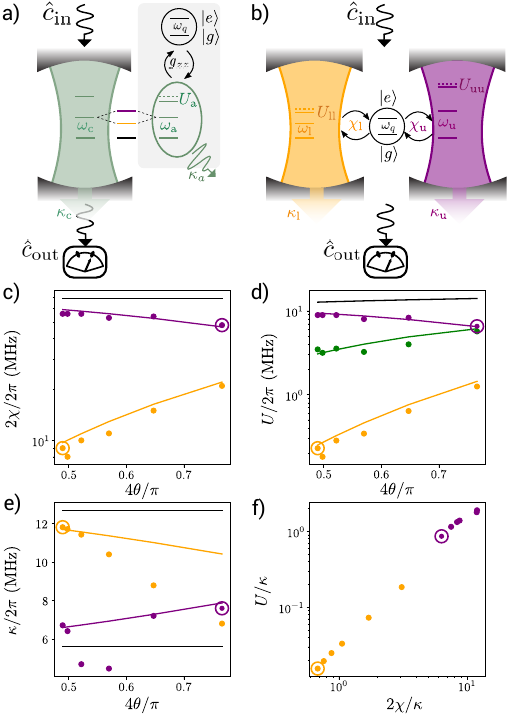}
\caption{a) Scheme of the setup. The cavity mode $\hat{c}$ of frequency $\omega_c$ is strongly coupled to an anharmonic ancilla mode of frequency $\omega_a$ and self-Kerr nonlinearity $U_a$. The ancilla is also coupled to the qubit via a non-perturbative cross-Kerr coupling of rate $g_{zz}$. To perform readout, we send a coherent signal on the input of the cavity mode $ \hat{c}_\mathrm{in}$ and measure the transmitted cavity output field $ \hat{c}_\mathrm{out}$. b) Representation of the system in terms of cavity-ancilla polariton modes. Lower and upper polariton modes have distinct frequencies $\omega_l$ and $\omega_u$, respectively, as well as different self-Kerr nonlinearities $U_l$ and $U_u$ inherited from the ancilla. Both polaritons are independently coupled to the qubit via cross-Kerr terms $\chi_l$ and $\chi_u$, which allows us to use these polariton modes as direct meters of the qubit states. The readout can be extracted from the same output field $ c_\mathrm{out} $ due to the polaritons leakage rate $\kappa_l$ and $\kappa_u$.
c)-e) Measurements (dots) and predictions (lines) for lower polariton $j=l$ (orange) and upper polariton $j=u$ (purple) as function of the hybridization angle $\theta$ of c) the non-perturbative qubit-polariton cross-Kerr $\chi_{j}$, d) the self-Kerr $U_{jj}$ and inter-polariton cross-Kerr $U_{ul}$ (green), and e) polariton decay rates $\kappa_j$. The predictions are calculated from the polariton model in Eqs. \eqref{eq:H_polaritons} and \eqref{eq:polariton_decay} using initial parameters $g_\mathrm{zz}$, $U_a$ and $\kappa_c$ and $\kappa_a$ and plotted as black lines in c), d) and e), respectively. f) Normalized self-Kerr polariton nonlinearity $U_{jj}/\kappa_j$ versus normalized qubit-polariton cross- Kerr coupling $2\chi_j/\kappa_j$ for lower and upper polaritons. c-f) The working points in the present work and in Ref. \cite{Dassonneville2020} are marked by a purple and orange circle, respectively. 
\label{fig:principe}}
\end{figure}

We use the same sample as in Ref. \cite{Dassonneville2020}. It consists of a 3D-cavity containing a transmon molecule which is made by coupling inductively and capacitively two nominally identical transmons \cite{Dumur2015, Dassonneville2020} (see \cref{fig:principe}a and \cref{App:sample_setup}). 
The device has three modes of interest, the harmonic TE$_{101}$ mode $\hat{c}$ of the rectangular $3$D Copper cavity with frequency $\omega_\mathrm{c}/2\pi = \SI{7.169}{GHz}$, and the two orthogonal modes of the transmon molecule (for more details see \cref{App:Molecule_Hamiltonian} and Ref. \cite{Dassonneville2020}): (i) a qubit mode, with parameters of standard transmons, protected from interaction with the 3D cavity mode thanks to its symmetric profile, and (ii) an ancilla mode, with weak ($\sim$ MHz) anharmonicity and strong interaction to the cavity $g_\mathrm{ac}$ due to its antisymmetric mode profile. Indeed, the cavity electrical field is aligned with the field of the ancilla mode while being orthogonal to the field of the qubit mode.

The ancilla is approximated as a weakly nonlinear mode $\hat{a}$ with frequency $\omega_{a}$ tunable by magnetic flux and self-Kerr rate $U_a$. Moreover, approximating the multilevel transmon as a qubit $\hat{\sigma}_z$, the total system Hamiltonian including cavity, ancilla, and qubit reads,
\begin{align}
\frac{\hat{H}}{\hbar} ={}& \frac{\omega_q}{2}\hat{\sigma}_z +\omega_a \hat{a}^\dag \hat{a} +\omega_c \hat{c}^\dag \hat{c} \notag\\ &- \frac{U_a}{2} \hat{a}^{\dag^2} \hat{a}^{^2} -g_{zz}\hat{\sigma}_z\hat{a}^\dag \hat{a}+g_{ac}(\hat{c}^\dag\hat{a}+\hat{a}^\dag\hat{c}).\label{FullHamqac}
\end{align}

The qubit $\hat{\sigma}_{\rm z}$ with frequency $\omega_q/2\pi =\SI{6.283}{GHz}$ and coherence times $T_2,~T_1 \simeq \SI{3}{\mu s}$ is coupled to the ancilla mode $\hat{a}$ via a \textit{non-perturbative} cross-Kerr coupling with rate $g_{\rm zz}$. The non-perturbative nature of this coupling allows to  maximize the speed, the single-shot
fidelity, and the QND properties of the readout, while minimizing the effect of unwanted decay
channels such as the Purcell effect \cite{Dassonneville2020}.

\subsection{Ancilla-cavity hybridization leading to polaritons}\label{polariton_transf}

To use this coupling for reading out the state of the qubit, we {strongly hybridize} the ancilla and the cavity by {setting} their detuning $\Delta_{ac} = \omega_a - \omega_c$ {to values comparable to or smaller than} their transverse coupling $g_{ac}/2\pi = \SI{295}{MHz}$.
At this operation point, $|\Delta_{ac}|\lesssim g_{ac}$, this hybridization leads to two new normal modes called upper and lower polariton modes, $\hat{c}_u$ and $\hat{c}_l$, for highest and lowest in frequency, respectively. They are a linear combination of ancilla and cavity fields (see \cref{fig:principe}.a-b). They are given by a rotation $\hat{c}_u =\cos(\theta)\hat{a}+\sin(\theta)\hat{c}$, and $\hat{c}_l =\cos(\theta)\hat{c}-\sin(\theta)\hat{a}$, where the cavity-ancilla hybridization angle reads $\tan(2\theta)= 2g_{ac}/ \Delta_{ac}$. At resonance ($\Delta_{ac}=0$, $\theta = \pi/4$), the two modes are completely hybridized into equal symmetric and antisymmetric superpositions while at large detuning ($|\Delta_{ac}| \gg g_{ac}$, $\theta \xrightarrow{} 0$), the two normal modes tend to approach the bare ancilla and cavity modes.

In terms of these polariton modes and using the rotating wave approximation, the total Hamiltonian takes the form
\begin{align} 
\frac{\hat{H}_{\rm p}}{\hbar} &= \frac{\omega_q}{2}\hat{\sigma}_z - \sum_{j=u,l} \chi_{j}\hat{c}^\dagger_j \hat{c}_j\hat{\sigma}_z  \notag \\
& + \sum_{j=u,l} (\omega_j \hat{c}_j^\dag \hat{c}_j - \frac{U_{jj}}{2} \hat{c}_j^{\dag^2} \hat{c}_j^{^2}) -U_{ul}\hat{c}^\dagger_l \hat{c}_l \hat{c}^\dagger_u \hat{c}_u ,
\label{eq:H_polaritons}
\end{align}
where $\omega_u= \sin^2(\theta) \omega_c + \cos^2(\theta)\omega_a + \sin(2 \theta) g_{ac} $ and $\omega_l = \cos^2(\theta)\omega_c + \sin^2(\theta)\omega_a- \sin(2 \theta)g_{ac}$ are the frequencies of the upper and lower polariton modes, respectively. Each polariton mode is in some proportion cavity-like and therefore can be probed in transmission and used for readout. Similarly, each polariton is also ancilla-like and thus inherits nonlinearities from the ancilla, notably the \textit{non-perturbative} cross-Kerr coupling to the qubit. The corresponding interaction strengths read $\chi_{u}= g_{zz}\cos^2(\theta)$ and $\chi_{l}= g_{zz}\sin^2(\theta)$, for the upper and lower polariton, respectively. Each polariton also inherits an anharmonicity from the ancilla given by $U_{ll} = \sin^4(\theta) U_a$ and $U_{uu} = \cos^4(\theta) U_a$. They also acquire a cross-anharmonicity or cross-Kerr interaction $U_{ul} = \sin^2(2\theta) U_a/2$, a coupling similar to the dispersive interaction that still occurs even beyond the dispersive regime \cite{Ansari2019}. Finally, the polaritons have effective decay rates given by a combination of the bare ancilla $\kappa_{a}$ and cavity $\kappa_{c}$ decay rates as (cf.~\cref{Append:bistability})
\begin{align}
   \kappa_u &=\kappa_c\sin^2({\theta})+\kappa_a\cos^2({\theta}), \notag \\
   \kappa_l&=\kappa_c\cos^2({\theta})+\kappa_a\sin^2({\theta}).
   \label{eq:polariton_decay}
\end{align}

\subsection{Tuning hybridization}

The ancilla can be tuned at discrete frequencies independently of the qubit and cavity. This is possible because the transmon molecule possesses two superconducting loops of different sizes with a high area ratio of 26 (see \cref{App:sample_setup} and \ref{App:Molecule_Hamiltonian}). Using one external coil, we can thus tune with a step precision of $\Phi_0/26$ the flux determining the ancilla frequency while the loop defining the qubit frequency still experiences an integer value of flux quantum $\Phi_0$. This allows to tune the hybridization conditions and thus the different parameters in Eqs.~\ref{eq:H_polaritons} and \ref{eq:polariton_decay}. We extracted these parameters (see Figs. \ref{fig:principe}.c-f) as function of the hybridization angle by measuring at different flux points. The non-perturbative cross-Kerr couplings $\chi_{l}$ and $\chi_{u}$ are well-fitted by a bare qubit-ancilla cross-Kerr coupling $g_{zz}/2\pi = \SI{34.5}{MHz}$ (\cref{fig:principe}.c). The polaritons self-Kerr and cross-Kerr couplings $U_{ll}$, $U_{uu}$, and $U_{lu}$ are well fitted with the polariton model with a bare ancilla anharmoncity $U_a/2\pi = \SI{13.5}{MHz}$ (\cref{fig:principe}.d). The polariton decay rates are only qualitatively fitted by the polariton model (\cref{fig:principe}.e) with bare cavity and bare ancilla decay rates $\kappa_c/2\pi = \SI{12.7}{MHz}$ and $\kappa_a/2\pi = \SI{5.6}{MHz}$. Discrepancies may be explained by the fact that the bare ancilla decay rate can vary with its frequency due to the presence of other losses like fluctuating two-level systems, or that there is residual parasitic transverse coupling between the ancilla and cavity with the qubit (see Ref.~\cite{Dassonneville2020}). 

Thanks to the hybridization tunability, we can set different regimes (\cref{fig:principe}.f) for the polariton meters to read out the qubit. In Ref. \cite{Dassonneville2020}, we focused on the linear response of the lower polariton which presents a small self-Kerr $U/\kappa = 0.017$ at the moderate drive $\overline{n} \approx 2\ll N_\mathrm{crit}$. This linear regime is obtained at the zero flux point ($\Phi/\Phi_0 = 0$) where the lower polariton is mostly cavity-like. In this work, we focus {on} a different regime where the anharmonicity is comparable to dissipation ($U\sim \kappa$). This is obtained for the upper polariton at $\Phi/\Phi_0 = 5$ when the ancilla is close to resonance to the cavity ($|\Delta_{ac}|\lesssim g_{ac}$). The hybridization is close to maximum, the upper polariton inherits from the ancilla an anharmonicity $U_{uu}/2\pi=\SI{6.5}{MHz}$ and has a loss decay $\kappa_u/2\pi=\SI{7.6}{MHz}$. At this working point, the cross-Kerr coupling to the qubit is the dominant parameter with $\chi_u/2\pi=\SI{24}{MHz}$. All other parameters for these two flux points are summarized in \cref{table}.

\begin{table}[h!]
	\begin{center}
		\begin{tabular}{c c c c c c}
	\hline
	$\Phi/\Phi_0 $ & ${\omega}_l/2\pi$	& $\chi_l/2\pi$ & $\kappa_l/2\pi$ & $U_\mathrm{ll}/2\pi $& $\bar{\theta}$  \\
		0 & 7.0335 {\rm GHz}  & $4.5 {\rm MHz}$ & \SI{11.8}{\mega\hertz} & \SI{0.2}{MHz} & \SI{0.384}{\radian} \\
	\end{tabular} 
	\\
	\begin{tabular}{c c c c c c}
	\hline
	$\Phi/\Phi_0 $ & ${\omega}_u/2\pi$	& $\chi_u/2\pi$ & $\kappa_u/2\pi$ & $U_\mathrm{uu}/2\pi $& $\bar{\theta}$  \\
		5 & 7.575 {\rm GHz}  & $24 {\rm MHz}$ & $7.6 {\rm MHz}$ & \SI{6.5}{MHz} & $\SI{0.602}{\radian}$ \\
		\hline 
	\end{tabular} 
		\caption{Effective polariton parameters at two flux points, $\Phi/\Phi_0=0$ corresponding to the working point in Ref~\cite{Dassonneville2020} and $\Phi/\Phi_0=5$ corresponding to the present working point. }
		\label{table}
	\end{center}  
\end{table}

\section{Qubit dependent polaritons response}
\label{sec:transmission}

\begin{figure*}
\includegraphics[width=0.9\textwidth]{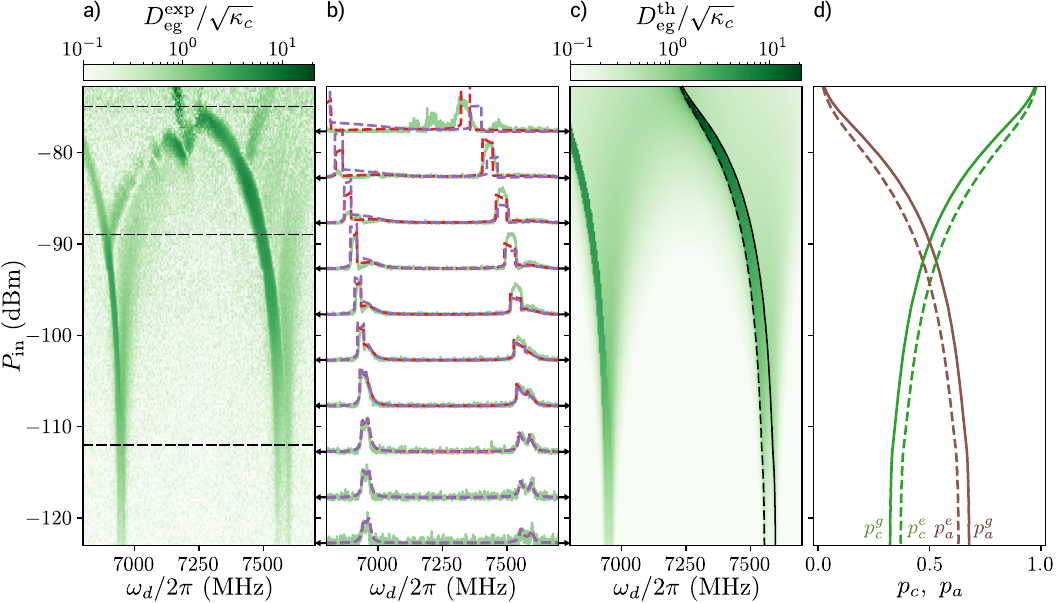}
\caption{ a) Measured mean distance $D_\mathrm{eg}^\mathrm{exp}$ between the two pointer states of the qubit as function of drive frequency $\omega_d$ and power $P_{\rm in}$. Horizontal dashed lines are guide to the eyes, indicating different regimes of the system. From bottom to top: {linear}, {nonlinear with self-Kerr}, higher order {nonlinearities}, and {strongly driven} bare cavity {regime}.
b) Cross-sections of $D_{eg}$ versus $\omega_d$ normalized by their experimental maximum value at fixed drive powers indicated by the arrows. In green: experimental data, in red: theoretical model in the ancilla-cavity basis (\cref{eq:polaritons_motion}), and in dashed purple: theoretical model in the polariton basis (\cref{nonlinear_polaritons}). c) Computed $D_\mathrm{eg}^\mathrm{th}$ using the model \cref{eq:polaritons_motion}.  
d) Decomposition of upper polariton into cavity {$p_{c}^\eta$} (green) and ancilla {$p_{a}^\eta$} (brown) as function of power $P_{\rm in}$ {when the qubit is in $\eta=g$ (solid lines) or $\eta=e$ (dashed lines)}. The proportions $p_{c}^\eta$ and $p_{a}^\eta$ are computed using model (\ref{eq:polaritons_motion}) and following the drive frequency-power line $\omega_d(P_{\rm in})$ {indicated by the solid and dashed lines in panel c), corresponding to the qubit states $\eta=g$ and $\eta=e$, respectively.} 
		 \label{fig:distance_map} }
\end{figure*}

{At time scales} much shorter than the qubit's lifetime $t\ll T_1$, the qubit state remains static in our setup and its main effect is to induce a qubit-dependent shift on the ancilla frequency as
\begin{align}
\omega_a\rightarrow {}&\bar{\omega}_a^{(\eta)}=\omega_a-g_{zz}\langle \hat{\sigma}_z\rangle_\eta,\label{shift}
\end{align}
where $\langle\hat{\sigma}_z\rangle_e=+1$ when the qubit is prepared in the excited state $\eta=e$ and $\langle\hat{\sigma}_z\rangle_g=-1$ when the qubit is in the ground state $\eta=g$. In terms of polaritons \cref{eq:H_polaritons}, this translates to a similar qubit-dependent shift on each polariton mode $j=u,l$ as
\begin{align}
\omega_j\rightarrow {}&\bar{\omega}_j^{(\eta)}=\omega_j-\chi_j\langle \hat{\sigma}_z\rangle_{\eta},
\end{align}
as well as to a change in the hybridization angle as $\theta\rightarrow \arctan\left[2g_{ac}/(\bar{\omega}_a^{(\eta)}-\omega_c)\right]/2$.

To observe this qubit-dependent shift on the polaritons experimentally, we drive the cavity with a coherent field $\langle \hat{c}_{\rm in }\rangle=i(\Omega_c/\sqrt{\kappa_c}) e^{-i\omega_d t}$ of frequency $\omega_d$ and amplitude $\Omega_c$. We then perform homodyne detection at the transmission output of the cavity $\langle  \hat{c}_{\rm out} \rangle_\eta=\sqrt{\kappa_c}\langle  \hat{c} \rangle_\eta$, which contains information of the qubit state $\eta=g,e$ and of the polaritons as $\langle  \hat{c}_{\rm out} \rangle_\eta =\sqrt{\kappa_c}(\sin(\theta) \langle  \hat{c}_u \rangle_\eta + \cos(\theta) \langle  \hat{c}_l \rangle_\eta)$. The experimental results are shown in Fig.~\ref{fig:distance_map}a, where we display the average pointer distance, 
\begin{align}
D_{eg} = \vert \langle \hat{c}_{\rm out}\rangle_{e} - \langle \hat{c}_{\rm out}\rangle_g \vert,
\end{align} 
as function of drive frequency $\omega_d$ and power $P_\mathrm{in}$, which relates to the amplitude as $\Omega_c= \sqrt{\kappa_c P_{\rm in}/\hbar\omega_d}$. For this, we integrate over a \SI{500}{ns} square readout pulse after preparing the qubit in its excited state ($\langle \hat{c}_{\rm out}\rangle_{e}$) or in its ground ($\langle \hat{c}_{\rm out}\rangle_{g}$) by applying a \SI{30}{ns} square $\pi$ pulse or not. The input power y-axis $P_\mathrm{in}$ is calibrated knowing the room temperature power and the attenuation in the input line. The colorbar of $D_{eg}$ is calibrated knowing the gain of the output line. For both calibrations of the input and output line, we assumed the calibrations to be flat in frequency, which is true in our frequency window up to \SI{\pm 1}{dB}.

As a first approximation, we consider both polaritons independent by neglecting their mutual coupling $U_{ul}$. It is thus possible to find a simple semi-classical model that properly describes our measurements {up to moderate powers}. When reaching a quasi steady-state $1/\kappa_c\ll t \ll T_1$, the polariton amplitudes are solutions of the nonlinear equations (cf.~\cref{Append:bistability}),
\begin{align}
    \langle \hat{c}_j\rangle_\eta = \frac{-i\Omega_j}{\kappa_j/2-i(\omega_d-\bar{\omega}_j^{(\eta)}+U_{jj} |\langle \hat{c}_j\rangle_\eta|^2)},\label{nonlinear_polaritons}
\end{align}
with effective polariton driving strengths $\Omega_u=\sin(\theta)\Omega_c$ and $\Omega_l=\cos(\theta)\Omega_c$.
For very weak driving, we can further neglect the term $U_{jj}$ in \cref{nonlinear_polaritons} and obtain the standard Lorentzians lineshapes. In this {linear} regime ($P_\mathrm{in} \lesssim \SI{-112}{dBm}$), we observe four peaks in $D_{eg}$ at the qubit-dependent polariton frequencies $\bar{\omega}_u^{(e)}=\omega_u + \chi_{u} = 2\pi\cdot\SI{7.599}{GHz}$, $\bar{\omega}_l^{(e)}=\omega_l + \chi_{l} = 2\pi\cdot\SI{6.963}{GHz}$, $\bar{\omega}_u^{(g)}=\omega_u - \chi_{u} = 2\pi\cdot\SI{7.552}{GHz}$, and $\bar{\omega}_l^{(g)}=\omega_l - \chi_{l} = 2\pi\cdot\SI{6.942}{GHz}$, {which can be resolved due to the large cross-Kerr shifts $2\chi_j\gtrsim \kappa_j$}. {The agreement between measurement and model (\ref{nonlinear_polaritons}) can be observed in Fig.~\ref{fig:distance_map}b), where we plot cross-sections of $D_{eg}$ for given input powers}. As the driving power increases, we need to consider the nonlinear term in the denominator of Eq.(\ref{nonlinear_polaritons}) and solve it self-consistently as a Duffing oscillator equation (cf.~\cref{Append:bistability}). As a result, the polariton frequencies are down-shifted due to their self-Kerr rate by an amount $\omega_j^{(\eta)}\rightarrow\bar{\omega}_j^{(\eta)}-U_{jj} |\langle \hat{c}_j\rangle_\eta|^2$ as shown in Figs.~\ref{fig:distance_map}a) and b). 
Above a critical value of driving strength $\Omega_{crit}$, the polaritons enter a bistability region, as can be hinted by the sharp wave-like shape {of the spectroscopic cross-sections above $P_{\rm in}\gtrsim -108$ dBm in} \cref{fig:distance_map}b. Contrary to the case of only one nonlinear resonator, here \cref{nonlinear_polaritons} presents two bistability zones close to each polariton frequency, similarly to two coupled nonlinear resonators \cite{Eichler2014}. This bistability region for the upper polariton is studied in the following \cref{sec:bistability}.

For the upper polariton, the effective model in \cref{nonlinear_polaritons} works well {up to $P_{\rm in}\lesssim -89$dBm} as indicated by the middle horizontal line in Fig.~\ref{fig:distance_map}a, but need to be refined at higher power. Indeed, 
describing a polariton by a fixed cavity-ancilla hybridization angle is no longer accurate at high power. As the driving power increases, the ancilla effectively modifies its frequency as $\bar{\omega}_a^{(\eta)}\rightarrow \bar{\omega}_a^{(\eta)} - U_a |\langle \hat{a}\rangle_\eta|^2$, which further modifies the hybridization condition as $\theta \rightarrow\arctan(2g_{ac}/ (\omega_a^{(\eta)}-U_a |\langle \hat{a}\rangle_\eta|^2-\omega_c) )/2$. 

To achieve {a more complete} description of the circuit, we consider the dynamics in the cavity-ancilla basis. As shown in \cref{Append:bistability}, we can still neglect quantum fluctuations, obtaining the following nonlinear system of equations for the quasi steady-state amplitudes:
\begin{align}
\label{eq:polaritons_motion}
    &[\kappa_c/2-i(\omega_d-\omega_c)]  \langle \hat{c}  \rangle_\eta + i g_{ac}  \langle \hat{a}  \rangle_\eta + i\Omega_c = 0,  \\
     &[\kappa_a/2-i(\omega_d-\bar{\omega}_a^{(\eta)}+U_a |\langle \hat{a} \rangle_\eta|^2)]\langle \hat{a}  \rangle_\eta + i g_{ac}  \langle \hat{c}  \rangle_\eta = 0.\nonumber
\end{align}
We map these equations to a standard Duffing oscillator polynomial equation of order 3 (see~\cref{Append:bistability}), which can have two stable solutions and one unstable solution (or vice versa) depending on the driving frequency $\omega_d/2\pi$ and amplitude $\Omega_c$. In Fig.~\ref{fig:distance_map}b) and {c) we show the theoretical prediction which agrees well with the cavity transmission measurements up to large powers}. The model \cref{nonlinear_polaritons}, which consider two independent polaritons, agrees well with the model \cref{eq:polaritons_motion}, which includes the coupling between polaritons, in a wide range of input power, up to $\gtrsim \SI{-89}{dBm}$. We conclude that the polariton modes are nearly uncoupled and thus can be used as independent meters for the qubit readout. Nevertheless, for calculation purposes, we will still perform many calculations using the ancilla-cavity basis in \cref{eq:polaritons_motion} rather than \cref{nonlinear_polaritons}.

From the numerical simulation of \cref{eq:polaritons_motion}, we also extract more information about the decomposition of the polariton modes in terms of cavity and ancilla components. In particular, we compute for both qubit states $\eta$ the cavity and ancilla population proportions, {$p_{c}^\eta = \frac{|\langle \hat{c}  \rangle_\eta|^2}{|\langle \hat{c}  \rangle_\eta|^2+|\langle \hat{a}  \rangle_\eta|^2}$} and {$p_{a}^\eta = \frac{|\langle \hat{a}  \rangle_\eta|^2}{|\langle \hat{c}  \rangle_\eta|^2+|\langle \hat{a}  \rangle_\eta|^2}$}, along the upper polariton branch $\omega_d$ as function of $P_\mathrm{in}$. The ancilla-cavity proportions depend on the qubit state as it shifts the ancilla frequency by $2g_{zz}$. At low power, the upper polariton is more ancilla-like than cavity-like while it is the opposite at large power. Effective ancilla-cavity resonance can be achieved for $|\expval{a}_\eta|^2 = (\bar{\omega}_a^{(\eta)} - \omega_c )/U_a$, where both polaritons become 50-50 ancilla-like and cavity-like. 

At even further input power, branches going to upward frequencies with input power start to appear in the measurement (\cref{fig:distance_map}.a). These features are not captured by our numerical model. We {believe} they are due to neglected higher nonlinear terms in the anharmonic ancilla mode, like the 6th order, which has an opposite sign to the 4th order self-Kerr term. The dynamics become more and more complex and is beyond the scope of this paper. However, we find it worthwhile to note that for a large enough driving power ($P_\mathrm{in} \geq \SI{-75}{dBm}$ in our case), the system reaches a regime where the polariton physics disappears and the qubit state can be readout close to the bare cavity frequency $\omega_\mathrm{c}/2\pi = \SI{7.169}{GHz}$. This behavior appears to be similar to the quantum-to-classical transition physics as in Ref. \cite{Reed2010} resulting in a destructive readout of the qubit state. At this high power, the ancilla is effectively far detuned from the cavity ($\theta \rightarrow 0$), the upper polariton is becoming more and more cavity-like until only the bare cavity is recovered (\cref{fig:distance_map}.d).

\begin{figure}
\includegraphics[width=8.6cm]{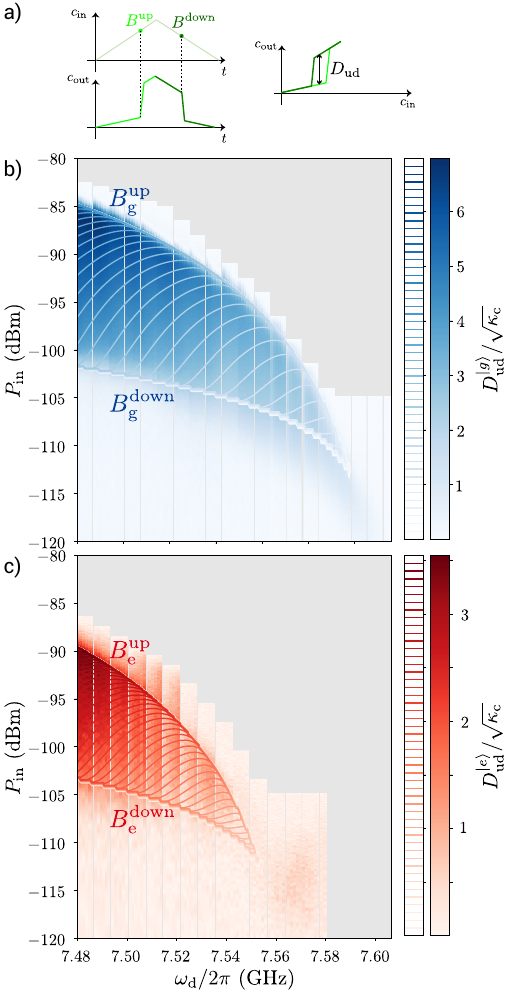}
	\caption{Bistability region for the upper polariton mode. (a) Schematic of the hysteretic behavior. Left: Pulse sequence $\Omega_c(t)$ for a ramp up and ramp down applied to cavity. This induces a bifurcation up (and down) in $\langle \hat{c}_{\rm out}\rangle^{\rm up}_\eta$ (and $\langle \hat{c}_{\rm out}\rangle^{\rm down}_\eta$), at different points $B^\mathrm{up}_\eta$ (and $B^\mathrm{down}_\eta$). Right: Hysteretic signal $D^\mathrm{ud}_{\eta}$ is reconstructed from the output signal during the ramp up and ramp down. (b)-(c) Measurement of the bistability hysteretic signal $D^\mathrm{ud}_\eta$ as a function of the input power $P_\mathrm{in}$ and frequency $\omega_d/2\pi$ when the qubit is prepared in $\eta=g$ (b) or in $\eta=e$ (c). The colormap corresponds to the experimental data and the contour lines correspond to the theory in \cref{eq:polaritons_motion}.
	   }
	\label{fig:bistability}
\end{figure}

\section{Qubit dependent upper polariton bistability regions}
\label{sec:bistability}

We now focus on frequencies around the upper polariton mode, and we study the bistability region in detail. As commented in the previous section, above a certain driving strength $\Omega_{crit}$ the nonlinear polariton can have two stable quasi steady states. When crossing a bistability zone (where two stable solutions coexist), the system presents an hysteric behavior. For a fixed drive frequency, and a ramp up in amplitude $\dot{\Omega}_c(t)>0$, the polariton bifurcates from a stable low amplitude output to a stable high amplitude output when crossing the point $B^{\rm up}_{\eta} (\omega_d, P_{\rm in})$ (see \cref{fig:bistability}a). Reciprocally, for a ramp down in amplitude $\dot{\Omega}_c(t)<0$, the polariton bifurcates from a high to a low amplitude output when crossing a different point $B^{\rm down}_\eta(\omega_d, P_{\rm in}) < B^{\rm up}_\eta(\omega_d, P_{\rm in})$. To characterize the bistability zone for the upper polariton, we measure the distance, 
\begin{align}
D^{\rm ud}_\eta=\vert \langle \hat{c}_{\rm out}\rangle_{\eta}^{\rm up}- \langle \hat{c}_{\rm out}\rangle_{\eta}^{\rm down}\vert,
\end{align}
which compares the output amplitude $\langle \hat{c}_{\rm out}\rangle^{\rm up}_\eta$ measured during a \SI{500}{ns} ramp up with the amplitude $\langle \hat{c}_{\rm out}\rangle^{\rm down}_\eta$ obtained during a \SI{500}{ns} ramp down. Both amplitudes are averaged over \SI{2000}{} realizations. The bistability region is identified by a non-zero hysteretic signal difference $D^{\rm ud}_\eta \neq 0$ since outside this region, we have $D^{\rm ud}_\eta = 0$. Remarkably, the bistability region of the upper polariton depends on the qubit state. 

In Figs.~\ref{fig:bistability}b) and c) we display the measured $D^{\rm ud}_\eta$ and the bistability regions, for the qubit prepared in its ground $\eta=g$ or excited state $\eta=e$, respectively. {From} the polaritons point of view, the qubit state shifts the upper polariton frequency by $2\chi_{u}$ and thus the bistability zone is also shifted by approximately $2\chi_{u}$. For both qubit states, the bistability zone and the hysteretic amplitude $D^{\rm ud}_\eta$ are well captured by \cref{eq:polaritons_motion}.

\section{Qubit state latching readout}
\label{sec:readout}

\begin{figure}
\includegraphics[width=8.6cm]{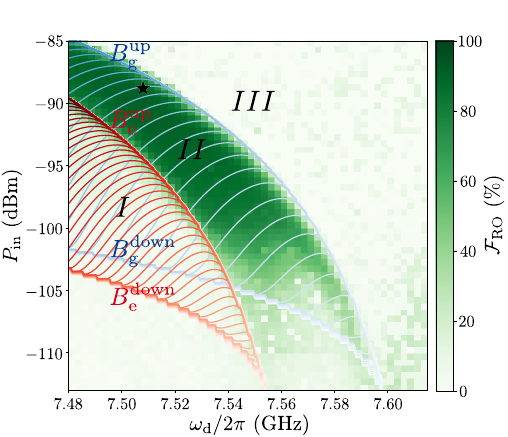}
	\caption{Readout fidelity as a function of power and frequency. Superimposed are the computed bistability zones for the upper polariton as shown in \cref{fig:bistability}. The point of maximum fidelity is indicated by a black star.  }
	\label{fig:fidelity}
\end{figure}
We measure the single-shot readout fidelity around the bistability zones, for the upper polariton, shown in \cref{fig:bistability}. The results of the readout fidelity as function of the frequency and power of the signal are shown in \cref{fig:fidelity}.
We see that for an amplitude below $B^{\rm up}_{e}$ (region I), the polariton does not bifurcate up for both qubit states. For amplitude above $B^{\rm up}_{g}$ (region III), the polariton bifurcates up for both qubit states.
Therefore, for regions I and III the readout fidelity is close to zero. This is not the case of region II, where we observe a very high fidelity. Here, at the beginning of the pulse, the input power is ramp up and crosses $B^{\rm up}_{e}$ but not $B^{\rm up}_{g}$. The upper polariton thus bifurcates up only if the qubit is in its excited state $\ket{e}$. During the pulse, the upper polariton will not bifurcate down as long as the input power does not fall below $B^{\rm down}_{g}$ or $B^{\rm down}_{e}$, even if the qubit relaxed to its ground state $\ket{g}$. The high amplitude output state is thus latched as long as input power is maintained above $B^{\rm down}_\eta$ even if the qubit relaxed. The readout is thus less sensitive to qubit relaxation error.
On the other hand, the readout is still sensitive to $\ket{g} \rightarrow \ket{e}$ transition error during measurement, for example due to thermal excitations.

Using heralding to mitigate state preparation errors, we measure a maximum readout fidelity of $\mathcal{F}_\mathrm{RO} = 1 - [P(e \vert g) + P(g \vert e)]/2 =\SI{98.6}{\%}$ for a \SI{500}{ns} square pulse.
Here, $P(\alpha \vert \beta)$ is the error probability to read state $\alpha$ when the system was prepared in state $\beta$. This fidelity is obtained at \SI{7.508}{GHz} drive frequency and \SI{-89}{dBm} input power (see \cref{fig:fidelity}). At this working point, the photon number of the readout mode is around $\overline{n}_g \sim 0$ and $\overline{n}_e \sim 9$ when the qubit is $\ket{g}$ and $\ket{e}$, respectively. The error probabilities $P(\alpha \vert \beta)$ are obtained by counting the statistic with thresholding from the readout histograms over $10^5$ repetitions where the qubit is prepared either in its ground or excited state.
We extract $P(e \vert g) = \SI{0.9 \pm 0.03}{\%}$ and $P(g \vert e) = \SI{1.9 \pm 0.05}{\%}$ where the uncertainty comes from the finite number of repetitions. 
During the readout integration time $t_\mathrm{int} = \SI{500}{ns}$, we would expect the qubit to have relaxed around $1-e^{-t_\mathrm{int}/2T_1} \simeq \SI{8}{\%}$ of the times. However, thanks to the latching bifurcation mechanism, this error is strongly reduced to around $1-e^{-t_\mathrm{b}/2T_1}$ where $t_\mathrm{b}$ is the time to bifurcate up to the high amplitude state, which is much smaller than the readout integration time $t_\mathrm{int}$. Assuming $t_\mathrm{b} \simeq \kappa_u^{-1} = \SI{21}{ns}$ as a rough estimate, we thus calculate \SI{0.3}{\%} error due to qubit relaxation before bifurcation within $P(g \vert e)$. We also estimate \SI{0.1}{\%} error due to overlaps between the two Gaussian corresponding to each qubit state, \SI{0.4}{\%} of heralding preparation error due to thermal excitations between the heralding and the readout pulse, \SI{1}{\%} of gate infidelity due to finite duration of $\pi$ pulse and finite coherence time. Within $P(e \vert g)$, we estimate \SI{0.1}{\%} error due to overlaps between the two Gaussian corresponding to each qubit state, \SI{0.4}{\%} of heralding preparation error due to thermal excitations between the heralding and the readout pulse and \SI{0.2}{\%} error due to thermal excitation during the readout. 
The remaining readout infidelity of \SI{0.15}{\%} (\SI{0.2}{\%} in $P(e \vert g)$ and \SI{0.1}{\%} in $P(g \vert e)$) could be attributed to wrong bifurcation event due to noise and uncertainty on the input power of the coherent drive. We believe that this wrong bifurcation event error to be small thanks to the large shift $2\chi_u/2\pi = \SI{48}{MHz}$ allowing to work with input power far enough from the $B_{\eta}^\mathrm{down/up}$ points. This error could be further suppressed in the future by optimizing the driving power and frequency working point and the readout pulse shape. 

The state of the qubit can thus be readout in a single-shot manner without any external quantum-limited amplifier. It is accomplished thanks to the qubit-state dependent bistability and enhanced by a latching mechanism. Here, this regime is achieved at low photon number (about 9) and differentiates strongly from the very high photon number like in Refs \cite{Reed2010,Gusenkova2021}.
Contrary to the case of in-situ JBA with transverse interaction to the qubit \cite{mallet_single-shot_2009, Schmitt2014, krantz_single-shot_2016}, here a large readout shift $\chi_u > \kappa_u, \text{ } U_{uu}$ is possible without suffering from Purcell decay thanks to the non-perturbative nature of the cross-Kerr coupling (see \cref{Append:comparison} for comparison). 
Using a two steps pulse \cite{Schmitt2014} or shelving techniques \cite{Schmitt2014, Elder2020, Chen2023} the errors due to wrong bifurcation events and relaxation before bifurcation may be even further reduced.
With a readily achievable qubit $T_1$ and $T_2$ ten times greater than the current sample, the false qubit preparation and the transitions before bifurcation errors could be suppressed to \SI{0.1}{\%}. Moreover, by adding an external quantum limited amplifier, the overlap error can be erased, and the input readout power further reduced, thus limiting backaction on the qubit. Taking into account the whole, readout fidelity beyond \SI{99.9}{\%} can be achieved.

\section{Conclusions and Outlook}

Using transmon molecule inside a cavity, we investigated transmon qubit readout based on polariton meters. In particular, we studied the nonlinear response of polariton modes to a strong drive leading to a hysteric bifurcation arising from the bistability of the upper polariton. Leveraging this effect, a latching-like readout with \SI{98.6}{\%} readout fidelity has been achieved without external quantum-limited amplification. This result is obtained in the mesoscopic regime when the polariton meters amplitude is limited to few photon number. Due to the non-perturbative cross-Kerr coupling between the qubit and the readout polariton modes, the qubit does not suffer from Purcell decay. A bifurcation amplifier of the qubit state with a large readout shift $2\chi>\kappa$ without deteriorating the qubit lifetime is thus possible and has been demonstrated. Such a readout with an in-situ amplification could reach quantum detection efficiency close to one by optimizing the techniques presented here. To this end, imperfections can be systematically studied and parameters recalibrated using recent QND detector tomography protocols \cite{Pereira_PRL_2022,pereira_parallel_2023,Rudinger2022}. Due to its flux tunability, our platform is also well-suited for a systematic study of qubit readout performances as function of $\chi$, $\kappa$, and $U$.

All the above properties make the transmon molecule inside a resonator a promising building block for implementing high-quality multiplexed readout in multi-qubit devices. Indeed, multiple transmon molecules could be simultaneously measured by coupling the dedicated resonators to a common transmission line, similarly as done in multi-qubit transmon platforms \cite{heinsoo_rapid_2018}. Although our scheme may be a priori slightly more complicated than the usual qubit-resonator-Purcell filter building blocks, which also involves three modes, the possibility of non-perturbative cross-Kerr couplings can strongly increase the speed and QND quality of the readout. This is particularly crucial for large-scale protocols with repetitive measurements \cite{krinner_realizing_2022,sundaresan_demonstrating_2023,acharya_suppressing_2023}. In particular, our scheme allows qubits to induce large readout shifts $2\chi>\kappa$ while still having large detuning to readout modes, thereby mitigating frequency crowding, enhancing speed, and suppressing non-QND errors originating from residual Purcell-like effects \cite{Dassonneville2020} or measurement-induced state transitions within and beyond the rotating-wave approximation \cite{PRL_Sank_2016, khezri2022measurementinduced}. In addition, having an intrinsic and in-situ amplification at each qubit can enable a large quantum detection efficiency, which can allow for a high signal-to-noise ratio even when strongly splitting the readout pulse to perform scalable multi-qubit measurements.

Finally, the transmon molecule circuit is also promising for other readout schemes using polaritons as an in-situ Josephson Parametric Dimer or JPA \cite{Eichler2014,Winkel2020}. This can be achieved through degenerate pumping either below the critical bistability points of each polariton or in between the frequencies of both polaritons. Implementing these strategies holds the potential to further enhance the quantum efficiency and performances of the readout process in future experiments.

\acknowledgments
The authors thank D. Vion and E. Dumur for fruitful discussions. R. D. acknowledges funding from CFM pour la recherche. V.M., C. M. and O. B. acknowledge support from ANR REQUIEM (ANR-17-CE24-0012-01) and ANR OCTAVES (ANR-21-CE47-0007-01). Work in Madrid is funded by the Spanish project PGC2018-094792-B-I00 (MCIU/AEI/FEDER, UE), CSIC Interdisciplinary Thematic Platform (PTI+) on Quantum Technologies (PTI-QTEP+), and Proyecto Sinergico CAM 2020 Y2020/TCS-6545 (NanoQuCo-CM). T.R. further acknowledges support from the Juan de la Cierva fellowship IJC2019-040260-I and from the Ramón y Cajal program RYC2021-032473-I, financed by MCIN/AEI/10.13039/501100011033 and the European Union NextGenerationEU/PRTR.

\appendix

\section{Sample and microwave wiring}
\label{App:sample_setup}

\begin{figure}
	\includegraphics[width=8.6cm]{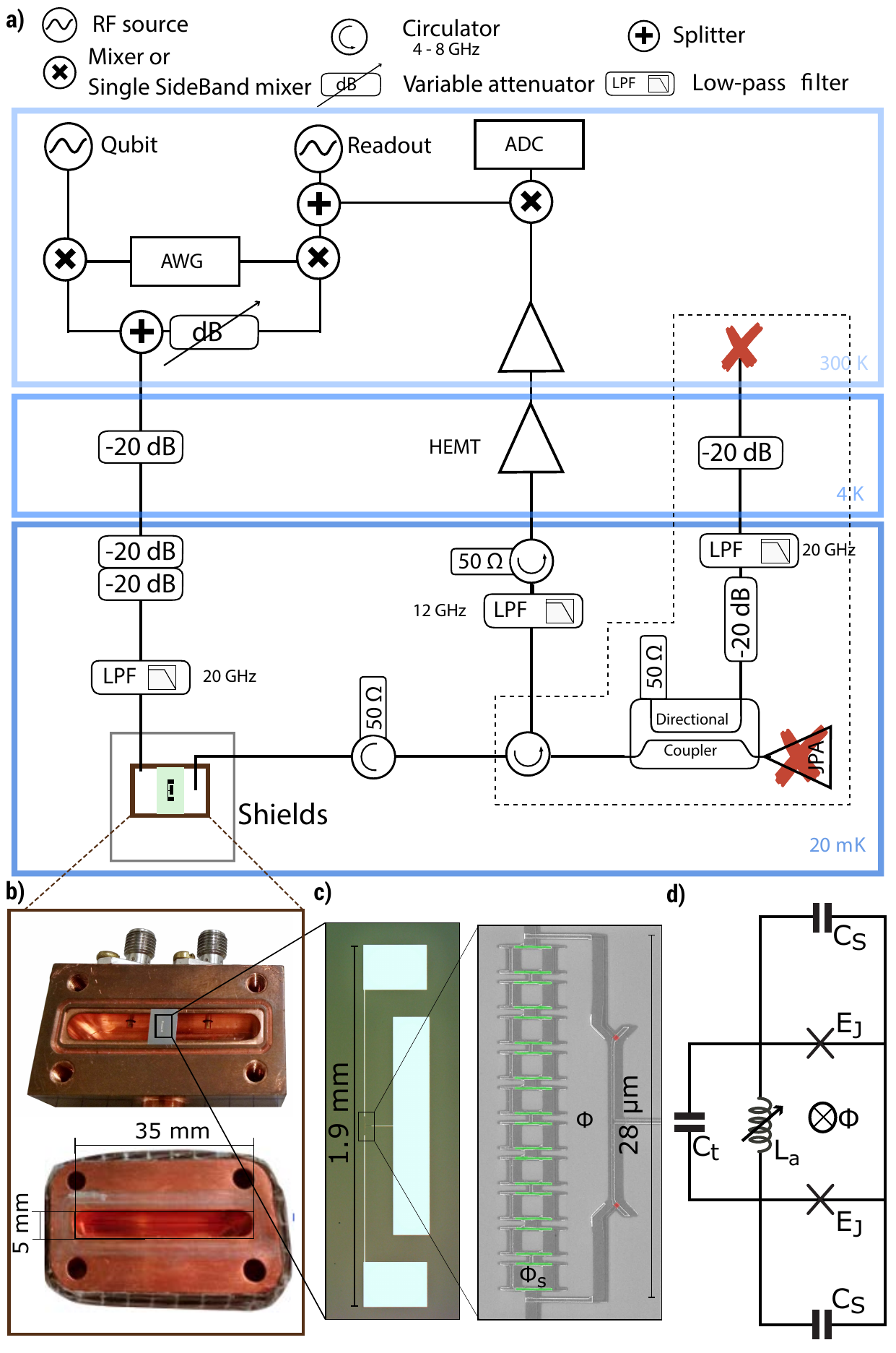}
	\caption{a) Schematic of the experimental setup. Even though a JPA is present, it is not pumped in the current work and so does not bring amplification. As our reported readout does not require external quantum-limited amplifier, the framed part with dash line can be removed in order to further optimize the output line and enhance the readout performances.  b) Picture of the two parts of the Copper-OFHC 3D cavity with the input-output pin connectors. The sample is placed at the center of the cavity. c) Optical microscope and SEM pictures of the transmon molecule sample. The Josephson junctions are highlighted in red. The SQUID Josephson junctions implementing the coupling inductance $L_a$ are highlighted in green. d) Lumped element circuit of the transmon molecule.}
	\label{Setup_figure}
\end{figure}

In this appendix, we describe the experimental setup. The device consists of an aluminium Josephson circuit realizing a transmon molecule and described by the lumped element circuit of Fig.~\ref{Setup_figure}d. The circuit is deposited on an intrinsic silicon wafer as shown in Fig.~\ref{Setup_figure}c and inserted in a 3D copper cavity (Fig.~\ref{Setup_figure}b). The transmon molecule Hamiltonian is described in the following appendix.

To measure the transmon molecule, we insert the silicon chip inside a 3D copper cavity with a volume \SI{24.5}{\milli\metre} x \SI{5}{\milli\metre} x \SI{35}{\milli\metre}{(cf.~Fig.~\ref{Setup_figure}b)}. The cavity mode considered is the fundamental TE$_{101}$ mode.

The schematic of the experimental set-up is shown in Fig.~\ref{Setup_figure}a. Qubit and readout pulses are sent through the same input line. The output line carries the readout signal to an HEMT through two circulators to prevent noisy microwave signal back to the cavity. The output line was wired to include a JPA which was turned off for the present study. Consequently, the transmitted signal passes through two circulators and a directional coupler before being reflected on the JPA. As the JPA was not pumped, no amplification were performed. 
Then the reflected signal passes again through the directional coupler and circulator before entering the first amplification stage with the HEMT amplifier. This not optimal output line wiring does not prevent to achieve high fidelity readout. However, the readout performances could be further improved by removing the JPA-related wiring that involve insertion and propagation losses. At room-temperature, the signal is down-converted to DC voltages via an IQ mixer and digitized at 1 GS/s using an ADC. Finally, the signal is digitally integrated.

\section{Transmon molecule Hamiltonian}
\label{App:Molecule_Hamiltonian}
In this appendix, we derive the circuit Hamiltonian of the transmon molecule. The molecule is realized by two identical transmon qubits with Josephson energy $E_J$ and capacitance $C_S$, coupled through a parallel LC-circuit with inductance $L_a$ and capacitance $C_t$ (see Fig.~\ref{Setup_figure}d). Here, $C_S$ represents the capacitance between an either small rectangular electrode and the central longer one, while $C_t$ represents the capacitance
between the two small rectangular electrodes. The coupling inductor
$L_a$ is implemented by a chain of 10 small SQUID loops of area $S_{\rm SQUID}$, which are tunable by an external flux $\Phi_{S}$ {(cf. Fig.~\ref{Setup_figure}c)}. The circuit also contains a large loop of enclosed area {$A$} that is approximately $r={A}/S_{\rm SQUID} \simeq 26$ times larger than the SQUIDs. {Consequently, the flux} $\Phi = r \Phi_S$
generates a circulating current passing through both $L_a$ and the two small Josephson junctions of the transmons. As already discussed in previous work~\cite{Dassonneville2020}, when the applied flux satisfies $\Phi =n\Phi_0$ (with $n$ an integer {and $\Phi_0$ the magnetic flux quantum}), the dynamics of {the system effectively behaves as a single transmon qubit with cross-Kerr coupling to a slightly anharmonic ancilla} mode, described by {the Hamiltonian}
\begin{align} 
\hat{H}_{\rm mol}={}& 4E_{C_q} \hat{n}_q^2 - 2E_J \cos(\hat{\varphi}_q)  \notag  \\
{}& + 4E_{C_a} \hat{n}_a^2 - 2E_J \left( \cos(\hat{\varphi}_a) - \frac{L_J}{L_a (n)} \hat{\varphi}_a^2\right) \notag \\ 
{}& - \frac{E_J}{2} \hat{\varphi}_q^2 \hat{\varphi}_a^2 + {\cal O}{^6}.
\label{Hamiltonian_Circuit} 
\end{align}
{Here, the phase average $\hat{\varphi}_q$ and phase difference $\hat{\varphi}_a$ between the two Josephson junctions describe the effective transmon qubit and the ancilla mode, respectively. Their conjugate charge number operators are denoted by $\hat{n}_{q}$ and $\hat{n}_{a}$. The charging energies of qubit and ancilla are given by $E_{C_{q}}=e^2/(2C_q)$ and $E_{C_{a}}=e^2/(2C_a)$, with effective capacitances $C_q=2C_S$ and $C_a=2(C_S+2C_t)$, respectively. We considered the system in the transmon regime, $E_J\gg E_{C_q}, E_{C_a}$, so that $\hat{\varphi}_q,\hat{\varphi}_a\ll 1$ and therefore expanded the coupling term between $\hat{\varphi}_q$ and $\hat{\varphi}_a$ up to fourth order in the phases}. In addition, $L_J= (\frac{\Phi_0}{2\pi})^2 \frac{1}{E_{J}}$ describes the Josephson inductance of each junction and $L_a (n)$ denotes the value of the coupling inductance for given magnetic flux $\Phi_{{S}} = \frac{n}{r} \Phi_0$. Importantly, the last term in Eq.~(\ref{Hamiltonian_Circuit}) {originates} the nonlinear {cross-Kerr} coupling between transmon qubit and ancilla as shown in the next subsection.

To express the Hamiltonian in the number representation, we exploit the analogy between the quadratic terms and the Hamiltonian of independent quantum harmonic oscillators with positions $\hat{x}_j=\hat{\varphi}_j$, momenta $\hat{p}_j=\hbar\hat{n}_j$, masses $m_j=\hbar^2/(8E_{C_j})$, and frequencies $\tilde{\omega}_j= \sqrt{8E_{J_j}E_{C_j}}/\hbar$, for qubit and ancilla ($j=q,a$), where we have defined the effective Josephson energies of qubit and ancilla as $E_{J_q}=2E_J$ and $E_{J_a}(n)=2E_J\left(1+\frac{2L_J}{L_a(n)}\right)$, respectively. With these identifications, we can use the known results from the quantization of the quantum harmonic oscillator and express the phase and number operators as
\begin{align}
    \hat{\varphi}_q ={}& \left(\frac{8E_{C_q}}{E_{J_q}}\right)^{1/4}\frac{(\hat{q}+\hat{q}^\dag)}{\sqrt{2}},\label{Op1}\\
    \hat{n}_q ={}& -i\left(\frac{E_{J_q}}{8E_{C_q}}\right)^{1/4}\frac{(\hat{q}-\hat{q}^\dag)}{\sqrt{2}},\\
    \hat{\varphi}_a ={}& \left(\frac{8E_{C_a}}{E_{J_a}(n)}\right)^{1/4}\frac{(\hat{a}+\hat{a}^\dag)}{\sqrt{2}},\\ 
    \hat{n}_a ={}& -i\left(\frac{E_{J_a}(n)}{8E_{C_a}}\right)^{1/4}\frac{(\hat{a}-\hat{a}^\dag)}{\sqrt{2}},\label{Op4}
\end{align}
where $\hat{q}$, $\hat{q}^\dag$ and $\hat{a}$, $\hat{a}^\dag$ are standard ladder operators for the qubit and ancilla modes, respectively. 

Replacing expressions (\ref{Op1})-(\ref{Op4}) into Eq.~(\ref{Hamiltonian_Circuit}), we diagonalize the quadratic terms of the circuit Hamiltonian, allowing us to interpret the qubit and ancilla modes as two coupled anharmonic oscillators described by  
\begin{align}
\frac{\hat{H}_{\rm mol}}{\hbar}={}& \tilde{\omega}_q \hat{q}^\dag \hat{q}+\frac{\alpha_q}{12}(\hat{q}+\hat{q}^\dag)^4+\tilde{\omega}_a \hat{a}^\dag \hat{a}\label{eq:hamiltonien}\\
{}&-\frac{U_a}{12}(\hat{a}+\hat{a}^\dag)^4-\frac{g_{zz}}{2}(\hat{q}+\hat{q}^\dag)^2(\hat{a}+\hat{a}^\dag)^2.\nonumber
\end{align}
Here, the anharmonicities of the qubit and ancilla are given by $\alpha_q = -E_{C_q}/\hbar$ and $U_a = (E_{C_a}/\hbar)(1+2\frac{L_J}{L_a(n)})^{-1}$, respectively, and $g_{zz} = \sqrt{\alpha_q U_a}$ is the strength of their cross-Kerr coupling.

Considering a large enough anharmonicity $|\alpha_q|$, we can approximate the transmon qubit by its two lowest states, so that it can be fully described by Pauli operators, $\hat{\sigma}_z = 2 \hat{q}^\dag \hat{q} -1$, and $\hat{\sigma}_- = \hat{q}$. Finally, adding the cavity mode $\hat{c}$, its coupling to the ancilla, and applying rotating wave approximation (RWA) to \cref{eq:hamiltonien}, we recover \cref{FullHamqac} of the main text.

\section{Open quantum system description and nonlinear steady-state dynamics}
\label{Append:bistability}

In this appendix, we present details of the theoretical models to describe our setup as a nonlinear open quantum system. We also indicate the approximations and the specific nonlinear Duffing equations that we use to interpret the measured data.

\subsection{Cavity-ancilla basis}

For a time-scale much shorter than the lifetime of the qubit, $t\ll T_1$, we can neglect the dynamics of the qubit, and therefore its only effect is to provide a static shift on the ancilla frequency depending on the qubit state. This is described by the system Hamiltonian $\hat{H}$ in Eq.~(\ref{FullHamqac}). If we additionally take into account the dynamics induced by a coherent drive of strength $\Omega_c$ on the cavity, $\hat{H}_d=\hbar\Omega_c(\hat{c}e^{i\omega_d t}+\hat{c}^\dag e^{-i\omega_{\rm d} t})$, the total Hamiltonian $\hat{H}_{\rm tot}=\hat{H}+\hat{H}_{\rm d}$ in the rotating frame with respect to the drive frequency $\omega_d$ reads
\begin{align}
    \frac{\hat{H}_{\rm tot}}{\hbar} ={}& - (\omega_d-\bar{\omega}_a^{(\eta)})\hat{a}^\dagger \hat{a} - \frac{U_a}{2} \hat{a}^{\dagger}\hat{a}^{\dagger} \hat{a} \hat{a}\label{Htot_rotating} \\
    {}&- (\omega_d-\omega_c) \hat{c}^\dagger \hat{c} + g_{ac}(\hat{a}^\dagger \hat{c} + \hat{c}^\dagger \hat{a}) + \Omega_c (\hat{c} + \hat{c}^\dagger)\nonumber,
\end{align}
where $\bar{\omega}_a^{(\eta)}=\omega_a-g_{zz}\langle\sigma_z\rangle_\eta$ is the qubit-dependent ancilla frequency. Considering the decay of the cavity $\kappa_c$ and ancilla $\kappa_a$, the master equation for the density operator of the system $\hat{\rho}(t)$ reads
\begin{align}
\dot{\hat{\rho}}  ={}& -i[\frac{\hat{H}_{\rm tot}}{\hbar},\hat{\rho}] + \kappa_c \mathcal{D}[\hat{c}]\hat{\rho} + \kappa_a \mathcal{D}[\hat{a}]\hat{\rho}.\label{Master_tot}
\end{align}

We consider a displacement transformation for the cavity and ancilla modes, $ \hat{a} = \alpha(t) + \delta \hat{a}$ and $ \hat{c} = \gamma(t) + \delta \hat{c}$, where $\delta \hat{a}$ and $\delta \hat{c}$ are quantum fluctuation operations around the classical mean displacements $\alpha(t)$ and $\gamma(t)$, respectively. The dynamics of the classical fields are given by the coupled nonlinear equations,
\begin{align}
    \dot{\alpha} ={}& -[\kappa_a/2 -i(\omega_d-\bar{\omega}_a^{(\eta)})]\alpha-ig_{ac}\gamma+iU_a|\alpha|^2\alpha,\nonumber\\
    \dot{\gamma} ={}& -[\kappa_c/2 -i(\omega_d-\omega_c)]\gamma-i\Omega_c-ig_{ac}\alpha,\label{classicalpart}
\end{align}
whereas the dynamics of the quantum fluctuations is described by the master equation for the new density operator $\hat{\rho}_F(t)$,
\begin{align}
\dot{\hat{\rho}}_F ={}& -i[\frac{\hat{H}_F}{\hbar},\hat{\rho}_F] + \kappa_c \mathcal{D}[\delta \hat{c}]\hat{\rho}_F + \kappa_a \mathcal{D}[\delta \hat{a}]\hat{\rho}_F.\label{fluctuation_master}
\end{align}
Here, the fluctuation Hamiltonian reads,
\begin{align}
    \frac{\hat{H}_F}{\hbar} ={}& - (\omega_d-\bar{\omega}_a^{(\eta)}+2U_a|\alpha|^2) \delta \hat{a}^\dagger \delta \hat{a} + g_{ac}(\delta \hat{a}^\dagger \delta \hat{c} + \delta \hat{c}^\dagger \delta \hat{a}) \notag \\ &- (\omega_d-\omega_c) \delta \hat{c}^\dagger \delta \hat{c} - \frac{U_a}{2} [\alpha^2 (\delta \hat{a}^\dagger)^2 + (\alpha^\ast)^2 (\delta \hat{a})^2] \notag \\ &- U_a[\alpha (\delta \hat{a}^\dagger)^2 \delta \hat{a} +\alpha^\ast \delta \hat{a}^\dagger \delta \hat{a}^2 ] - \frac{U_a}{2} (\delta \hat{a}^\dagger)^2(\delta \hat{a})^2.\label{fluctuation_Hamiltonian}
\end{align}
Note that the dynamics in Eqs.~(\ref{classicalpart})-(\ref{fluctuation_Hamiltonian}) is equivalent as the original one in Eqs.~(\ref{Htot_rotating})-(\ref{Master_tot}), but the former equations are more convenient when quantum fluctuations are small, which is the case when driving the system with a strong coherent drive on the cavity $\Omega_c$.

The steady state solution for the classical displacements in Eqs.~(\ref{classicalpart}) is obtained by first solving the following nonlinear algebraic Duffing oscillator for the ancilla,
\begin{align}
    [A-i(B+C|\alpha_{\rm ss}|^2)]\alpha_{\rm ss}=D,\label{Duffing1}
\end{align}
and then obtaining the solution for the cavity displacement as
\begin{align}
    \gamma_{\rm ss}=\frac{-i(\Omega_c+g_{ac}\alpha_{\rm ss})}{\kappa_c/2-i(\omega_d-\omega_c)}.\label{Duffing3}
\end{align}
Notice that in Eq.~(\ref{Duffing1}) the parameters read $A= \kappa_a/2+g_{ac}{\rm Re}(z)$, $B=\omega_d-\bar{\omega}_a^{(\eta)}-g_{ac}{\rm Im}(z)$, $C = U_a$, $D=-z\Omega_c$, and $z= g_{ac}/[\kappa_c/2-i(\omega_d-\omega_c)]$. This Eq.~(\ref{Duffing1}) is more conveniently solved in terms of a third order polynomial, 
\begin{align}
    C^2 x^3 +2BC x^2 +(A^2+B^2)x-D=0,\label{Duffing2}
\end{align}
for $x=|\alpha_{\rm ss}|^2$. In this work, we numerically solve for Eq.~(\ref{Duffing2}) and then determine the full ancilla displacements via $\alpha_{\rm ss}=D/[A-i(B+Cx)]$ and Eq.~(\ref{Duffing3}).

We checked numerically that in steady state $t\gg 1/\kappa_c$, the quantum fluctuations are always small with respect to the classical displacement, $|\langle \delta \hat{a} \rangle|\ll |\alpha_{\rm ss}|$ and $|\langle \delta\hat{c}\rangle|\ll |\gamma_{\rm ss}|$, and therefore it is a good approximation to completely neglect them, namely $\langle \hat{a} \rangle_\eta\approx \alpha_{\rm ss}$, and $\langle \hat{c} \rangle_\eta\approx \gamma_{\rm ss}$ as given by the solution of Eqs.~(\ref{Duffing1})-(\ref{Duffing2}). Under these approximations, is that Eq.~(\ref{eq:polaritons_motion}) of the main text is valid.

In Sec.~\ref{sec:bistability}, when discussing the bi-stability of the upper polariton, we calculate the stable solutions after a ramp up and ramp down of the system by the solutions of Eq.~(\ref{Duffing2}) with lowest and largest amplitude, respectively. This matches very well the measured data as indicated in Figs.~\ref{fig:bistability}b)-c).

\begin{table}[h!]
	\begin{center}
		\begin{tabular}{|c| c| c  | c | c|}
	\hline
	References  & present work	&  \cite{mallet_single-shot_2009} &  \cite{Schmitt2014} & \cite{krantz_single-shot_2016} \\
	\hline
		$\chi/2\pi$ ($\si{MHz}$) & 24 & 2.1 &  1.7 & 3.6\\
		$\kappa/2\pi$ ($\si{MHz}$) & 7.6 & 9.4 & 3.1 & 1.3\\
		$ U/2\pi$ ($\si{MHz}$) & 6.5  & (0.02)  & 0.3 & 0.03\\
		$N_\mathrm{crit}$ & 0.2 & 90.5 &  2.0 & 8.3 \\
		$\mathcal{F}_\mathrm{RO}$ ($\%$) & 98.6 & 92 & 98.3 & 90.8 \\
		$t_\mathrm{int}$ (\si{ns}) & 500 & 965 & 2025 & 600\\  
		$T_1$ ($\si{\mu s}$) & 3.3 & 0.5  & 2.5 & 4.1\\
		\hline 
	\end{tabular} 
		\caption{Readout system parameters for different references using a Josephson bifurcation amplifier or a Josephson parametric oscillator.}
		\label{table_chi_kappa_U}
	\end{center}  
\end{table} 
\subsection{Polariton basis}

Starting from Eqs.~(\ref{Htot_rotating})-(\ref{Master_tot}) and doing a polariton transformation $\hat{c}_u =\cos(\theta)\hat{a}+\sin(\theta)\hat{c}$, and $\hat{c}_l =\cos(\theta)\hat{c}-\sin(\theta)\hat{a}$, as indicated in Sec.~\ref{polariton_transf}, we can describe the system by the master equation,
\begin{align}
\dot{\hat{\rho}}' ={}& -i[\frac{\hat{H}^{\rm p}_{\rm tot}}{\hbar},\hat{\rho}'] + \sum_j\kappa_j \mathcal{D}[\delta \hat{c}_j]\hat{\rho}',
\end{align}
where the total Hamiltonian for the polariton modes $\hat{c}_j$, considering the coherent drive in the rotating frame with frequency $\omega_d$, reads 
\begin{align}
\frac{\hat{H}^{\rm p}_{\rm tot}}{\hbar}={}& \frac{\omega_q}{2}\hat{\sigma}_z - \sum_{j=u,l} \chi_{j}\hat{c}^\dagger_j \hat{c}_j\hat{\sigma}_z+ \sum_{j=u,l} (\omega_j \hat{c}_j^\dag \hat{c}_j - \frac{U_{jj}}{2} \hat{c}_j^{\dag^2} \hat{c}_j^{^2})\notag\\ 
{}&+\sum_{j=u,l}\Omega_j(\hat{c}_j+\hat{c}_j^\dag)-U_{ul}\hat{c}^\dagger_l \hat{c}_l \hat{c}^\dagger_u \hat{c}_u,
\label{eq:H_polaritons2}  
\end{align}
Here, the effective polariton driving strengths are given as in the main text as $\Omega_u=\sin(\theta)\Omega_c$ and $\Omega_l=\cos(\theta)\Omega_c$. We can also apply a displacement transformation on the polariton operators, $\hat{c}_j=\gamma_j(t)+\delta \hat{c}_j$ and obtain an equivalent description, similar as done above for the cavity-ancilla basis. The classical displacements $\gamma_j(t)$ are described by the coupled nonlinear equations,
\begin{align}
    \dot{\gamma}_j ={}& - [\kappa_j/2-i(\omega_d-\omega_j^{(\eta)})]\gamma_j -i\Omega_j +i U_{jj} |\gamma_j|^2 \gamma_j\nonumber\\ 
    {}&+ i U_{ul} |\gamma_{-j}|^2 \gamma_j.
\end{align}
The quantum fluctuations are described by a master equation,
\begin{align}
\dot{\hat{\rho}}_F' = -i[\hat{H}_{\rm p}^F, \hat{\rho}_F' ] + \kappa_j \mathcal{D}[\delta\hat{c}_j]\hat{\rho}_F'.
\end{align}
The Hamiltonian for the polariton quantum fluctuations $\hat{H}_{\rm p}^F$ can be more compactly written using the identification $\delta \hat{c}_u\rightarrow \delta \hat{c}_+$ and $\delta \hat{c}_l\rightarrow \delta \hat{c}_-$ as
\begin{align}
\hat{H}^F_{\rm p} ={}& \sum_{j=\pm}\left[-(\omega_d-\bar{\omega}_{j}^{(\eta)}+2U_{jj}|\gamma_j|^2+\frac{U_{ul}}{2}|\gamma_{-j}|^2)\delta \hat{c}_j^\dagger \delta \hat{c}_j\right.\notag\\ 
{}&- U_{jj}\gamma_j^\ast \delta \hat{c}_j^\dagger (\delta \hat{c}_j)^2 - U_{jj} \gamma_j (\delta \hat{c}_j^\dagger)^2 \delta \hat{c}_j - \frac{U_{jj}}{2} (\gamma_j)^2 (\delta \hat{c}_j^\dagger)^2\notag\\ 
{}&-\left. \frac{U_{jj}}{2} (\gamma_j^\ast)^2 (\delta \hat{c}_j)^2 - \frac{U_{jj}}{2} (\delta \hat{c}_j^\dagger)^2(\delta \hat{c}_j)^2\right]\notag\\
{}&-U_{ul}\sum_{j=\pm}\left[\gamma_j(\gamma_{-j}^\ast)\delta \hat{c}_j^\dagger \delta \hat{c}_{-j} + \gamma_{-j}^\ast \delta \hat{c}_j^\dagger \delta \hat{c}_j \delta \hat{c}_{-j}\right.\notag\\ 
{}&+ \gamma_{-j} \delta \hat{c}_{-j}^\dagger \delta \hat{c}_j^\dagger \delta \hat{c}_{j} + \frac{\gamma_j}{2} \gamma_{-j} \delta \hat{c}_j^\dagger \delta \hat{c}_{-j}^\dagger\notag\\ 
{}&+\left. \frac{1}{2} \gamma_j^\ast \gamma_{-j}^\ast \delta \hat{c}_j \delta \hat{c}_{-j}+ \frac{1}{2}\delta \hat{c}_j^\dagger \delta \hat{c}_{-j}^\dagger \delta \hat{c}_j \delta \hat{c}_{-j} \right].
\end{align}

Similarly, as for the cavity-ancilla basis, we checked numerically that quantum fluctuations are always small in our problem, $|\langle \delta \hat{c}_j \rangle|\ll |\gamma_j|$, and therefore we neglect them for simplicity $\langle \hat{c}_j \rangle_\eta\approx \gamma_j$.

In addition, we found that neglecting the coupling between polaritons $U_{ul}$ is a good approximation to our data up to moderately large driving amplitudes $\Omega_c$. This further simplifies the analysis of the nonlinear polariton meters of our setup, as they can be well modeled as independent modes. The quasi steady state amplitudes of each polariton $\langle \hat{c}_j \rangle_\eta$ are solutions of standard Duffing oscillator equations of the form,
\begin{align}
    [\kappa_j/2-i(\omega_d-\bar{\omega}_j^{(\eta)}+U_{jj} |\langle \hat{c}_j \rangle_\eta|^2)]\langle \hat{c}_j \rangle_\eta=-i\Omega_j.
\end{align}
This is Eq.~(\ref{nonlinear_polaritons}) of the main text, which can be put in the same form of Eq.~(\ref{Duffing2}) with $x_j=|\langle \hat{c}_j \rangle_\eta|^2$, $A_j=\kappa_j/2$, $B_j=\omega_d-\bar{\omega}_j^{(\eta)}$, $C_j=U_{jj}$, and $D_j=-i\Omega_j$. In this way we solve self-consistently for the occupation of each polariton mode $x_j=|\langle \hat{c}_j \rangle_\eta|^2$ first, and then we obtain the full solution for each mode independently as $\langle \hat{c}_j \rangle_\eta=D_j/[A_j-i(B_j+C_jx_j)]$.

\section{Comparison with other bifurcation readout}
\label{Append:comparison}

In \cref{table_chi_kappa_U} are summarized the $\chi$, $\kappa$, $U$, $N_\mathrm{crit}$, $\mathcal{F}_\mathrm{RO}$, $t_\mathrm{int}$ and $T_1$ values for our present work and for different references using a Josephson Bifurcation Amplifier or a Josephson Parametric Oscillator with dispersive interaction with a qubit. While $\kappa$ is in the same order of magnitude, our cavity pull $\chi$ is much larger than in other references thanks to the non-perturbative nature of the cross-Kerr coupling. Moreover, JBA is usually used in the case $U \ll \kappa$ meaning a large number of photons is required to attain the bistability zone while in our case, $U \lesssim \kappa$, photon number close to unity is sufficient to attain the bistability zone.

\input{main.bbl}

\end{document}

%% file: main.bbl
%